# Gaussian fluctuation of the diffusion exponent of virus capsid in a living cell nucleus

Yuichi Itto

*Science Division, Center for General Education, Aichi Institute of Technology,*

*Aichi 470-0392, Japan*

**Abstract.** In their work [Proc. Natl. Acad. Sci. USA 112 (2015) E5725], Bosse et al. experimentally showed that virus capsid exhibits not only normal diffusion but also anomalous diffusion in nucleus of a living cell. There, it was found that the distribution of fluctuations of the diffusion exponent characterizing them takes the Gaussian form, which is, quite remarkably, the same form for two different types of the virus. This suggests high robustness of such fluctuations. Here, the statistical property of local fluctuations of the diffusion exponent of the virus capsid in the nucleus is studied. A maximum-entropy-principle approach (originally proposed for a different virus in a different cell) is applied for obtaining the fluctuation distribution of the exponent. Largeness of the number of blocks identified with local areas of interchromatin corrals is also examined based on the experimental data. It is shown that the Gaussian distribution of the local fluctuations can be derived, in accordance with the above form. In addition, it is quantified how the fluctuation distribution on a long time scale is different from the Gaussian distribution.

# 1. Introduction

There is growing interest in viruses from the viewpoint of physics. Examples include the mechanical property (such as elasticity) of viruses, self-assembly of viruses as a thermodynamic process (phase transition), the use of viruses in nanotechnology, etc. (see Refs. [1-3], for example). They reveal intriguing aspects of the physical properties of viruses and related phenomena.

In their recent work [4] (see also Ref. [5]), Bosse et al. have experimentally studied the diffusion property of herpesviruses in nuclei of living PtK2 cells by making use of the technique of single particle tracking. (Here, the herpesvirus is an enveloped virus particle, whereas the PtK2 cell is a kidney epithelial cell.) This virus contains a protein shell, which is referred to as capsid. In the experiment, the cells were infected with pseudorabies virus (i.e., suid herpesvirus 1) or herpes simplex virus 1, the capsid of which is labeled with a fluorescent protein. The nucleus is organized by chromatin (i.e., chromosomal substance), by which interchromatin compartments called corrals are made. Such an organization offers a "sponge-like" architecture of the nucleus, in which these compartments represent for example enclosed pores. Due to this, large macromolecules are trapped in the corrals. In fact, the fluorescent virus capsids have been found to diffuse inside the corrals in the following situation. During virus infection, chromatin structure becomes more porous, making the corral size increase, and simultaneously, viral replication compartments (i.e., sites of viral DNA replication) form in the nucleus.

In order to characterize the diffusion property, the mean square displacement of the virus capsid, which is expressed here by $\overline{x^2}$, has been evaluated by analyzing tracks of the capsids in the corrals. Then, it has been found to scale as

$$\overline{x^2} \sim t^\alpha \tag{1}$$

for elapsed time, $t$. The experimental result shows not only normal diffusion but also anomalous diffusion: in the former, $\alpha = 1$, whereas $\alpha \neq 1$ in the latter. A remarkable observation there is that $\alpha$, which is termed here the diffusion exponent, fluctuates for a wide range from subdiffusion, i.e., $0 < \alpha < 1$, to superdiffusion, i.e., $\alpha > 1$. Since the virus capsids are uniformly distributed over the nucleus [4], this fluctuation seems to occur in different corrals on a large spatial scale. Then, the corresponding distribution of fluctuations has been presented. Surprisingly, as can be seen in Fig. 2 (C) in Ref. [4], this distribution takes the Gaussian form (see also Ref. [5]).

It is here worth briefly mentioning recent advances in experimental studies of anomalous diffusion based on the technique of single particle tracking (see Refs. [6,7] for its theory and foundations, for example). It has been shown in Ref. [8] that lipid granules in living cells exhibit subdiffusion. In Ref. [9], it has been observed that a particle such as vesicle in a pathogen shows superdiffusion. In addition, the impressiveness of the technique with molecular simulation has been discussed in Ref. [10].

It should be also noted that the phenomenon of exponent fluctuations itself is equally nontrivial as well as anomalous diffusion [11,12] widely discussed in the literature.

We describe the Gaussian distribution observed in the experiment as follows:

$$f(\alpha) \sim \exp\left[-\frac{(\alpha - \alpha_0)^2}{2\sigma^2}\right], \tag{2}$$

where $\alpha_0 = 0.85$ and $\sigma = 0.24$ are the mean value and the standard deviation of $\alpha$, respectively [4]. In particular, it is seen to behave as $f(0) \sim 0$.

Here is the important feature concerning the distribution in Eq. (2). That is, the Gaussian distribution is *robust* in the sense [4,5] that it takes the same form for two different herpesviruses (i.e., the pseudorabies virus and the herpes simplex virus 1). Thus, the experimental result presents an evidence for high robustness of the observed exponent fluctuations.

Now, for a different virus in cytoplasm of a different cell, over which the diffusion exponent fluctuates depending on localized areas [13], a theoretical approach has been developed in recent works [14-16] for describing the statistical property of the local fluctuations of the diffusion exponent. There, a statistical distribution of the local fluctuations has been proposed based on experimental data. Then, it has been shown that the proposed distribution can be derived by the maximum entropy principle for such exponent fluctuations.

As mentioned earlier, exponent fluctuations for the virus capsids seem to occur in different corrals over the nucleus. This indicates that the diffusion exponent, $\alpha$, in Eq. (1) fluctuates depending on the corrals: that is, the local fluctuations of $\alpha$. Therefore, this naturally motivates us to examine if the above approach can offer the Gaussian distribution in Eq. (2). This issue has an obvious importance for understanding the robustness of the Gaussian fluctuation (in the above-mentioned sense) deeper. In addition, such an understanding sheds new light on structure of the nucleus as a medium for diffusion of the virus capsid. In fact, it has been shown, in recent experimental studies (e.g., [17-24]) of diffusion of proteins concerning other viruses and other proteins in cells, that knowledge about local fluctuations yields nontrivial feature of cellular structure such as its heterogeneity (as a medium for diffusion), although fluctuating quantity discussed there is not the diffusion exponent but the diffusion coefficient.

In the present paper, we study the statistical property of the local fluctuations of the diffusion exponent of a herpesvirus capsid in nucleus of a living PtK2 cell. We regard the region of corrals over the nucleus as a medium for diffusion of the capsids of both pseudorabies virus and herpes simplex virus 1. Then, imaginarily dividing it into many small blocks, we introduce entropy associated with such fluctuations. Based on the experimental data of subdiffusion, the case of which is an explicitly tractable one in the data, we also examine how large the number of these blocks is. We show that, in accordance with the form in Eq. (2), the Gaussian distribution of the fluctuations can be derived by the maximum-entropy-principle approach [14-16], in which the statistical behavior of the exponent at its second order is considered to play an informative role. In addition, we quantitatively show how the fluctuation distribution after long duration of time is different from the Gaussian distribution in Eq. (2).

**2. Entropy associated with the fluctuations of the diffusion exponent**
  **and largeness of the number of blocks**

We start our discussion with considering diffusion of the virus capsid in the nucleus. As a first step, taking the presence of chromatin into account, we regard the region of the corrals over the nucleus as a medium for diffusion of the capsids of both the pseudorabies virus and the herpes simplex virus 1. Then, we imaginarily divide this medium into many small blocks, in which the virus capsid exhibits anomalous diffusion as well as normal diffusion. Therefore, the diffusion exponent, $\alpha$, in Eq. (1) fluctuates depending on these local blocks. It is noted [4] that $\overline{x^2}$ in Eq. (1) is evaluated for the elapsed time smaller than that taken for estimation of the size of corral. Accordingly, a local block is identified with a certain area of the corral. (We will quantitatively discuss this point later.) Although the diffusion exponent may slowly vary over a period of time, we assume that $\alpha$ is approximately constant, here.

We introduce the entropy associated with exponent fluctuations. Our situation is that no information is available about how the exponent locally fluctuates over the region of the corrals. So, we consider a collection formed by constructing the local blocks, which is equivalent to the medium at the statistical level of the fluctuations. As we shall see below, the medium is, in fact, expected to contain many blocks, and accordingly, there are a large number of distinct collections in terms of the local fluctuations. Suppose that the medium consists of the blocks with a set of different values of discrete exponents, $\{\alpha_i\}_i$. For this medium, let $N$ and $n_{\alpha_i}$ be the number of total blocks and the number of blocks with the $i$ th exponent, $\alpha_i$, respectively. It is clear that the relation, $\sum_i n_{\alpha_i} = N$, is fulfilled. Then, for average taken in the analysis of $\overline{x^2}$, by which the exponent in a given area of the corral is determined, it is considered that tracks in other areas of the corral as well as other corrals are not included in it. This naturally suggests that the blocks are independent each other with respect to the exponent. Therefore, the total number of distinct collections, $G$, is given by $G = N! / \prod_i n_{\alpha_i}!$. With it, we then introduce the entropy as follows:

$$S = \frac{\ln G}{N}. \tag{3}$$

Next, we present a discussion about largeness of the number of blocks by evaluating the volume of the region of the corrals. Observation turns out to indicate that $S$ in Eq. (3) takes the form of the Shannon entropy.

In Ref. [5], the volume of the nucleus is assumed to be given by about $570\,\mu m^3$. Accordingly, we consider that the volume of the region of the corrals is smaller than this volume, since chromatin is present in the nucleus as mentioned earlier. We further stress the following points. As mentioned in the Introduction, during virus infection, it has been observed that the corral size increases since chromatin structure becomes more porous. Simultaneously, it has also been found that viral replication compartments form in the nucleus. These compartments may occupy the volume of the nucleus less compared to the corrals for a period of time, over which the Gaussian fluctuation in Eq. (2) is realized [4]. Therefore, in the present paper, we suppose that the volume of the region of the corrals is not so small but still large enough.

Now, for a given local block in the medium, we treat it as a cubic block, the side of which has the length of the value of $\sqrt{x^2}$ at large elapsed time. This treatment (originally performed for a localized area of cytoplasm in Ref. [25]) enables us to estimate the volume of the cubic block. As the value of $\alpha$ in Eq. (1), we take, as explicitly tractable cases, the following two values presented in Fig. 2 (B) in Ref. [4]: $\alpha = 0.961$ for the pseudorabies virus and $\alpha = 0.918$ for the herpes simplex virus 1. Regarding these exponents, it should be noted [4] that each exponent is obtained by average of the mean square displacement for all tracks over the nucleus. Then, the diffusion coefficient, which is necessary for estimating $\overline{x^2}$ in our discussion, presented there is also obtained in this way. Such a diffusion coefficient as well as the diffusion exponent may be different from those to be used in $\overline{x^2}$ in Eq. (1), but they are sufficient in the present discussion since the difference is expected to be small [4]. Therefore, for the elapsed time $0.36\,\text{s}$ [4], the volumes of the cubic block are estimated as follows: $0.022\,\mu\text{m}^3$ for the pseudorabies virus and $0.013\,\mu\text{m}^3$ for the herpes simplex virus 1. Recalling that the local block is identified with a certain area of the corral, these volumes are, in fact, smaller than that of the corral [5], as expected. In each case, if the total number of blocks with the above-estimated volume in the medium is, for example, 100, then it follows that $n_{0.961} = 100$ in the former, while $n_{0.918} = 100$ in the latter, for each of which $N > 100$, provided that no additional blocks with the corresponding exponent are supposed to be present. So, in this example, the total volumes of the blocks with each of the above volumes are calculated to be $2.2\,\mu\text{m}^3$ and $1.3\,\mu\text{m}^3$, respectively. Thus, the volume of the region of the corrals is seen to be large enough compared to these volumes. From this observation, it is therefore implied that $n_{0.961}$ and $n_{0.918}$ are actually larger than the above values and the numbers of blocks with other exponents are also large, since the volume of the medium should be equal to that of the region of the corrals.

3. **Derivation of the Gaussian fluctuation by the maximum entropy principle**

From the above discussion about the blocks, it seems that the entropy in Eq. (3) can be evaluated by the use of the Stirling approximation [i.e., $\ln(M!) \cong M \ln M - M$ for large $M$]. This leads to the following form of the Shannon entropy:

$S \cong -\sum_i f_{\alpha_i} \ln f_{\alpha_i}$, where $f_{\alpha_i} = n_{\alpha_i}/N$ is the probability of finding $\alpha_i$ in a given local block of the medium. Therefore, we write its continuum limit as

$$S[f] = -\int d\alpha\, f(\alpha) \ln f(\alpha), \tag{4}$$

where we are using the same symbol $f(\alpha)$ for the probability density as that in Eq. (2), but it may not lead to confusion.

Let us show that the Gaussian distribution can be derived by the maximum entropy principle with the Shannon entropy in Eq. (4).

Recall that there is no information about how the exponent locally fluctuates over the region of the corrals. It is natural in such a situation to impose the constraint on the average of $\alpha$ in the principle:

$$\int d\alpha\, f(\alpha)\, \alpha = m. \tag{5}$$

Then, as stressed earlier, viral replication compartments form during virus infection. The fluctuation distribution in the vicinity of the above average may slightly change due to these compartments, but we consider the fluctuation distribution that is stable under such a formation over a period of time. Accordingly, additional constraint seems to be required in order to take into account this stability, since a class of the fluctuation distributions to be realized is limited by it with keeping the above average. To do so as simple as possible, we impose the constraint on the statistical behavior of $\alpha$ at its second order, i.e., the size of the fluctuations of $\alpha^2$. Therefore, if additional information is available on such a statistical behavior, then the average of $\alpha^2$, i.e., the second moment, should be further constrained:

$$\int d\alpha\, f(\alpha)\, \alpha^2 = s^2. \qquad (6)$$

It is noticed from Eqs. (5) and (6) that the variance of $\alpha$ is automatically fixed, reflecting the above-mentioned stability. Thus, together with another constraint on the normalization of $f(\alpha)$, the corresponding maximum entropy principle reads

$$\delta_f \left\{ S[f] - \kappa \left( \int d\alpha\, f(\alpha) - 1 \right) + \lambda \left( \int d\alpha\, f(\alpha)\, \alpha - m \right) \right.$$

$$\left. - \mu \left( \int d\alpha\, f(\alpha)\, \alpha^2 - s^2 \right) \right\} = 0, \qquad (7)$$

where $\{\kappa, \lambda, \mu\}$ is the set of the Lagrange multipliers associated with the constraints, and $\delta_f$ stands for the variation with respect to $f(\alpha)$. Note that we have imposed the following two conditions in Eq. (7):

$$\lim_{\alpha \to \infty} f(\alpha) = 0, \qquad f(0) < f(\alpha^*), \qquad (8)$$

where $\alpha^*$ is a certain fixed positive value of $\alpha$. The limit, $\alpha \to \infty$, should be interpreted as the large-$\alpha$ limit, i.e., the limit $\alpha$ going to a large but finite value. The first condition in Eq. (8) is nothing but the requirement of physically plausible behavior of $f(\alpha)$ due to the fact that the region of the corrals is finite. Then, the second one comes from the following natural considerations. The virus capsid tends to reach the membrane of the nucleus for egress (i.e., a budding process), implying the existence of many blocks with a typical large value of $\alpha$. This tendency might seem to indicate that such a value could be very large. This is however seen to lead to contradiction regarding the above first condition, and therefore we take the value as $\alpha^*$. Due to the time dependence in Eq. (1), the tendency then may become strong in the case when the number of blocks with the exponent $\alpha = \alpha^*$ is larger than that of blocks with the exponent $\alpha = 0$, which yields the second condition. It turns out that the conditions in Eq. (8) require $\lambda$ and $\mu$ to be positive Lagrange multipliers.

Consequently, the stationary solution of Eq. (7) is given by

$$\hat{f}(\alpha) \propto \exp(\lambda \alpha - \mu \alpha^2). \tag{9}$$

Clearly, this distribution can be written in the following form, $\hat{f}(\alpha) \propto \exp\{-\mu[\alpha - \lambda/(2\mu)]^2\}$, and accordingly, becomes identical to the distribution in Eq. (2) after the following identifications:

$$\lambda = \frac{\alpha_0}{\sigma^2}, \qquad \mu = \frac{1}{2\sigma^2}. \tag{10}$$

Thus, we see that the maximum-entropy-principle approach can give rise to the Gaussian distribution of the local fluctuations, which is precisely $f(\alpha)$ in Eq. (2).

It is anticipated from the present results that the average of $\alpha^2$ in Eq. (6) can be informative in the approach.

## 4. Comment on the difference between the fluctuation distributions on a long time scale

So far, we have discussed the situation that the virus capsids are uniformly distributed over the nucleus, in which the Gaussian fluctuation in Eq. (2) is realized. In contrast to this situation, after long duration of time, it has been observed in the experiment [4] that the virus capsids are marginalized around the membrane of the nucleus in the case of the pseudorabies virus, which is nontrivial [5]. Accordingly, the fluctuation distribution to be observed on such a long time scale seems to be different from Eq. (2). So, it is of interest to quantify the difference between the Gaussian distribution in Eq. (2) and such a fluctuation distribution, since it may be possible to measure how the marginalization makes the Gaussian distribution approach the fluctuation distribution to be observed. Therefore, we wish to make a comment on this issue.

Suppose that Gaussian form similar to Eq. (2) still holds for the fluctuation distribution to be observed, $\tilde{f}(\alpha)$, but the associated mean value and standard deviation can differ from those in Eq. (2) and are denoted by $\tilde{\alpha}_0$ and $\tilde{\sigma}$, respectively. (This is supported by the raw data of the experiment [4], although the data is not shown here.) Then, we quantify the difference between $f(\alpha)$ and $\tilde{f}(\alpha)$ by the distance between them. Such a distance may be supplied by the symmetric Kullback-Leibler divergence:

$$D[\tilde{f}, f] = K[\tilde{f} \| f] + K[f \| \tilde{f}]. \qquad (11)$$

Here, $K[\tilde{f} \| f]$ describes the Kullback-Leibler relative entropy [26] and is given in the present case by

$$K[\tilde{f} \| f] = \int d\alpha \, \tilde{f}(\alpha) \ln \frac{\tilde{f}(\alpha)}{f(\alpha)}, \qquad (12)$$

which is positive semidefinite and vanishes if and only if $\tilde{f} = f$. Thus, we obtain

$$K[\tilde{f} \| f] = \frac{1}{2\sigma^2} \left\{ (\tilde{\alpha}_0 - \alpha_0)^2 + \tilde{\sigma}^2 \right\} - \ln \frac{\tilde{\sigma}}{\sigma} - \frac{1}{2}, \qquad (13)$$

$$K[f \| \tilde{f}] = \frac{1}{2\tilde{\sigma}^2} \left\{ (\tilde{\alpha}_0 - \alpha_0)^2 + \sigma^2 \right\} + \ln \frac{\tilde{\sigma}}{\sigma} - \frac{1}{2}, \qquad (14)$$

leading to

$$D[\tilde{f}, f] = \frac{1}{2} \left( \frac{1}{\tilde{\sigma}^2} + \frac{1}{\sigma^2} \right) (\tilde{\alpha}_0 - \alpha_0)^2 + \frac{1}{2} \left( \frac{\tilde{\sigma}}{\sigma} - \frac{\sigma}{\tilde{\sigma}} \right)^2. \qquad (15)$$

As an example of interest, we consider the case when $\tilde{\sigma} = \sigma$. In this case, we have

$$D[\tilde{f}, f] = \frac{1}{\sigma^2} (\tilde{\alpha}_0 - \alpha_0)^2, \qquad (16)$$

showing that $\tilde{f}(\alpha)$ deviates from $f(\alpha)$ in a monotonic way with respect to $\tilde{\alpha}_0$. This fact also illustrates that if the fluctuation distribution in the course of the marginalization remains the Gaussian form, then the distribution approaches $\tilde{f}(\alpha)$ in this way. Thus, we quantitatively see how the fluctuation distribution to be observed on a long time scale differs from the Gaussian distribution in Eq. (2).

**5. Concluding remarks**

We have studied the statistical property of the local fluctuations of the diffusion exponent of a herpesvirus capsid over the region of corrals in nucleus of a living PtK2 cell. Regarding the region as a medium for diffusion of the capsids of both pseudorabies virus and herpes simplex virus 1, and imaginarily dividing it into many small blocks, which are identified with local areas of the corrals, we have introduced the entropy associated with the fluctuations. Treating the local block as the cubic one, we have also examined largeness of the number of blocks. Based on the maximum entropy principle, in which the second moment of the exponent plays an informative role, we have derived the Gaussian form of the fluctuation distribution in consistent with the experimental observations. This highlights a possibility of applying the present approach to the diffusion-exponent fluctuations observed in the experiment. In addition, we have quantitatively discussed how the fluctuation distribution to be observed on a long time scale is different from the Gaussian distribution in Eq. (2).

In the present study, we have focused our attention on the fluctuations of the diffusion exponent. It may be natural to suppose that the diffusion coefficient of the virus capsid also fluctuates in the nucleus. So, it could be expected by the entropic approach that, if the diffusion coefficient satisfies similar constraints and conditions discussed for the diffusion exponent, then its statistical distribution is of the Gaussian form.

In addition, in a recent work [27], a unimodal distribution of the diffusion-exponent fluctuations of a viral protein in a cell nucleus has experimentally been observed.

Therefore, it is of interest to examine application of the present approach for such a distribution.

**Acknowledgements**

The author would like to thank J. B. Bosse for providing him with the raw data of the experiment in Ref. [4] and for valuable discussions. He also would like to acknowledge the warm hospitality of Heinrich Pette Institute, Leibniz Institute for Experimental Virology.

*Note added*. In a recent paper [28], an issue regarding the continuum limit of the entropy [14-16] has been carefully examined. The discussion given there is seen to be applicable to the entropy in Eq. (4).

**References**

[1] P.G. Stockley, R. Twarock (Eds.), Emerging Topics in Physical Virology,
    Imperial College Press, London, 2010.

[2] W.H. Roos, R. Bruinsma, G.J.L. Wuite, Nat. Phys. 6 (2010) 733.

[3] M.G. Mateu (Ed.), Structure and Physics of Viruses,
    Springer, Dordrecht, 2013.

[4] J.B. Bosse, I.B. Hogue, M. Feric, S.Y. Thiberge, B. Sodeik, C.P. Brangwynne, L.W. Enquist, Proc. Natl. Acad. Sci. USA 112 (2015) E5725.

[5] J.B. Bosse, L.W. Enquist, Nucleus 7 (2016) 13.

[6] F. Höfling, T. Franosch, Rep. Prog. Phys. 76 (2013) 046602.

[7] C. Manzo, M.F. Garcia-Parajo, Rep. Prog. Phys. 78 (2015) 124601.

[8] J.-H. Jeon, V. Tejedor, S. Burov, E. Barkai, C. Selhuber-Unkel, K. Berg-Sørensen, L. Oddershede, R. Metzler, Phys. Rev. Lett. 106 (2011) 048103.

[9] J.F. Reverey, J.-H. Jeon, H. Bao, M. Leippe, R. Metzler, C. Selhuber-Unkel, Sci. Rep. 5 (2015) 11690.

[10] J.-H. Jeon, M. Javanainen, H. Martinez-Seara, R. Metzler, I. Vattulainen, Phys. Rev. X 6 (2016) 021006.

[11] J.-P. Bouchaud, A. Georges, Phys. Rep. 195 (1990) 127.

[12] R. Metzler, J.-H. Jeon, A.G. Cherstvy, E. Barkai, Phys. Chem. Chem. Phys. 16 (2014) 24128.

[13] G. Seisenberger, M.U. Ried, T. Endreß, H. Büning, M. Hallek, C. Bräuchle, Science 294 (2001) 1929.

[14] Y. Itto, J. Biol. Phys. 38 (2012) 673.

[15] Y. Itto, Phys. Lett. A 378 (2014) 3037.

[16] Y. Itto, in: Atta-ur-Rahman, M. Iqbal Choudhary (Eds.), Frontiers in Anti-Infective Drug Discovery, Vol. 5, Bentham Science Publishers, Sharjah, 2017.

[17] S. Manley, J.M. Gillette, G.H. Patterson, H. Shroff, H.F. Hess, E. Betzig, J. Lippincott-Schwartz, Nat. Methods 5 (2008) 155.

[18] N. Dross, C. Spriet, M. Zwerger, G. Müller, W. Waldeck, J. Langowski, PLoS ONE 4 (2009) e5041.

[19] T. Kühn, T.O. Ihalainen, J. Hyväluoma, N. Dross, S.F. Willman, J. Langowski, M. Vihinen-Ranta, J. Timonen, PLoS ONE 6 (2011) e22962.

[20] S. Bakshi, B.P. Bratton, J.C. Weisshaar, Biophys. J. 101 (2011) 2535.

[21] E. Adu-Gyamfi, M.A. Digman, E. Gratton, R.V. Stahelin, Biophys. J. 102 (2012) 2517.

[22] N. Hoze, D. Nair, E. Hosy, C. Sieben, S. Manley, A. Herrmann, J.-B. Sibarita, D. Choquet, D. Holcman, Proc. Natl. Acad. Sci. USA 109 (2012) 17052.

[23] B.-C. Chen et al., Science 346 (2014) 1257998.

[24] H. Anton, N. Taha, E. Boutant, L. Richert, H. Khatter, B. Klaholz, P. Rondé, E. Réal, H. de Rocquigny, Y. Mély, PLoS ONE 10 (2015) e0116921.

[25] Y. Itto, Physica A 462 (2016) 522.

[26] S. Kullback, Information Theory and Statistics, Dover, New York, 1997.

[27] B.L. Rice, R.J. Kaddis, M.S. Stake, T.L. Lochmann, L.J. Parent, Front. Microbiol. 6 (2015) 925.

[28] Y. Itto, Open Conf. Proc. J. 9 (2018) 1.